# Plasmonic orbital angular momentum manipulation through light control

X.-C. Yuan[1*], Z. J. Hu[1], G. H. Yuan[2] and Z. Shen[1]

[1] Institute of Modern Optics, Key Laboratory of Optical Information Science & Technology, Ministry of Education of China, Nankai University, Tianjin, 300071, China

[2] School of Electrical & Electronic Engineering, Nanyang Technological University, Nanyang Avenue, 639798, Singapore

*Corresponding author: xcyuan@nankai.edu.cn

**Abstract:** Plasmonic vortices (PV) excited by a highly focused radially polarized optical vortex (RPOV) beam on a metal surface are investigated experimentally and theoretically. The proposed method reveals a direct phase singularity and orbital angular momentum (OAM) transfer from an incident structured beam to its counterpart in surface plasmon with dynamic, reconfigurable and high-efficiency advantages. The plasmonic field pattern, phase distributions, Poynting vector and focusing efficiency of PV are studied in detail. Experimental verification further shows that nanoparticles can be confined and manipulated within the region of PV and orbital rotation speed of the trapped particles is altered dynamically by changing the topological charge of the incident light.

Surface plasmon polaritons (SPP) are electromagnetic surface modes trapped at metal/dielectric interface because of their strong interaction with free electrons of the metal [1]. The most attractive characteristics of SPP are high localization and remarkable field enhancement in nature, allowing them to be controlled and manipulated in nanometer scales [2-4]. To achieve similar functionalities of traditional electrical circuits on plasmonic chips, for example guiding, switching and multiplexing, effective excitation and modulation of SPP with the help of nanostructures are required and reported recently, where fundamental plasmonic generators are employed to produce various diffractive SPP patterns on the metal/dielectric interfaces, originating from focusing and interference of SPP [5-14]. Efficient manipulation of SPP is not only concerned with the plasmonic focusing properties in terms of spot size, shape and magnitude, but also and more importantly the phase distribution and wavefront behaviors of the SPP with additional capabilities in manipulation. Orbital angular momentum (OAM) of optical vortices (OV) has widely been studied in the optics regime for applications in optical manipulation, microscopic image processing and information encoding and decoding in free-space communications [15-22]. Recently, it was reported that plasmonic vortices (PVs) can be produced by surface relief spiral grooves or a plasmonic vortex lens (PVL) under illumination of circularly polarized beam, where the spiral grooves or PVL served as gratings for wavevector matching and generation of plasmonic phase singularity [23-25]. Similar to OV in the detection of spiral wavefronts, the PVs can be employed to detect the SPP properties in the near field, for example the handedness of incident circularly polarized beam. As reported in [26, 27], a spin-based plasmonic effect in coaxial nanoapertures was used to couple the intrinsic spin angular momentum of the incident radiation to the extrinsic OAM of the SPP via spin-orbit interaction. It is noted, however, that although a surface relief spiral structure surrounding a nanoaperture can confer the necessary phase conditions as mentioned above, the requirement of center alignment and fabrication of complex structures in the nanometer scale are found difficult in practice and it becomes bottleneck in many applications.

In combination with respective merits of axially symmetric polarized beam, in this paper, we propose an all-optical method for focusing SPP to generate PV beyond the diffraction limit, which retains the phase singularity of incident OV beam and OAM conferred by the helical beam structure. We have been proposing generation and manipulation of SPP all-optically without surface relief structures. The technique enables excitation of SPP happened at the uniform metal/dielectric interface where incident light beams satisfied the incident angle and polarization conditions. It is noted that the method benefits from structureless excitation of SPP with dynamic configuration. In our early work, the SPP were generated successfully for excitation of fluorescent molecules and wide-field high resolution microscopy [28]. Recently, we extended the method to the generation of plasmonic linear phase singularity with a linearly polarized OV beam [29]. This leads to a challenging idea that the PV should be produced in a similar way but using a radially polarized OV beam (RPOV) and its OAM behavior in manipulation can experimentally be verified as its counterpart in the optical regime. Although examples of optical OAM manipulation by OV beams have widely been reported in terms of angular momentum conversion between the beam and trapped particles [30, 31], experimental verification of OAM manipulation in a plasmonic way represents a great challenge because of its near field nature. In addition, the all-optical excitation technique provides advantageous freedom in terms of dynamic configuration.

**Results**

**Experimental generation and characterization of PV.**

Figure 1 shows the schematic diagram of our experimental setup. A linearly polarized incident beam of wavelength 633 nm (He-Ne laser) was converted to circular polarization after passing through a quarter waveplate. Subsequently, the circularly polarized beam illuminated on a spiral phase plate encoded with topological charge $l$ to generate circularly polarized OV. If the handedness of the circular polarization is in the opposite direction of the spiral phase, a circularly polarized OV of topological charge $l$-1 will be produced, while a circularly polarized OV of topological charge $l$+1 will be yielded if the handedness of the circular polarization and spiral phase are consistent. After that, the radial (azimuthal) component can be extracted by a radial-type (azimuthal-type) polarization analyzer (PA). A polarization rotator (PR) consisting of two half-wave plates can enable direct switching between radial polarization and azimuthal polarization. In this way, the radially polarized optical vortices (RPOV) of topological charge $l$-1 or $l$+1 can effectively be generated and utilized for excitation of SPP on top surface of the metallic layer. The modulated wavefront is then transferred by a telescope to the back aperture of an oil-immersion objective lens (Olympus 100× NA 1.49) and normally focused onto a 50-nm-thick Au film ($\varepsilon_2$=-9.514+1.218i at 633 nm wavelength) deposited on the standard cover glass ($n$=1.515) along the +$z$ direction. The polarization states after some key optical elements are sketchily given in the square with yellow background colour for better understanding. The gap between the objective lens and the cover slip is filled with index matching oil. The finite NA value of the objective lens imposes a limit (~79.6°) to the range of angles, part of which corresponding to the surface plasmon resonance (SPR) condition lies in between the maximum angle of the lens and the critical angle for total internal reflection.

In high NA condition, RPOV emanating from the objective lens and converging towards the geometric focus will give rise to diffraction-limited spots containing a large spectrum of wavevectors limited by NA of the lens. The focusing property of the resulting RPOV is firstly examined using Richard-Wolf vectorial diffraction theory [32, 33]. Since polarization and phase are related to each other, the electric field of this phase modulated RPOV after highly focusing has radial ($E_r$), azimuthal ($E_\varphi$) and longitudinal ($E_z$) components, whereas $E_r$ and $E_z$ are p-polarized and $E_\varphi$ is s-polarized with respect to the interface. Therefore, when the focused RPOV imprints on the Au film, $E_\varphi$ will be attenuated due to its relatively low transmission efficiency over the entire angle ranges while $E_r$ and $E_z$ have a remarkable transmission near the SPR angle $\theta_{sp}$=44.5° within which the incident energy can efficiently be coupled to SPP through the Au film. In the following, we concentrate on discussion of the longitudinal plasmonic field $E_z$ because it is prominently enhanced in comparison with the transverse components and plays a dominant role in characterizing the electric field distributions on the metal film. There is no direct pervious propagating beam near the beam center owing to the central dark region of incident RPOV. The SPP excited along the diameter of the focused incident OV act as secondary sources and propagate inwardly if the Au film is located before the focal plane of the objective lens. Hence the central plasmonic field is attributed to the superposition of the initial SPP based on Huygens-Fresnel principle, and it can mathematically be expressed as

$$E_{1z}(r,\varphi,z) = \int_0^{2\pi} E_{1z} e^{il\theta} e^{-k_z \cdot z} e^{-ik_{sp}\bar{e}_r \cdot (\bar{r}-\bar{R}_0)} d\theta = E_{1z} e^{-k_z \cdot z} e^{ik_{sp}R_0} e^{il\varphi} J_l(k_{sp}r) \qquad (1)$$

where $E_{1z}$ is the initial amplitude of SPP along the designed radius ($R_0$) of incident RPOV, ($r$, $\varphi$) denote the observation point in cylindrical coordinate, $k_{sp}$ and $k_z$ are the transversal and longitudinal wave vectors, respectively. Here we have neglected the propagation loss of SPP because the imaginary part of $k_{sp}$=(1.061+0.000173i)$k_0$ at the wavelength 633 nm is very small. It is seen from Eq. (1) that the plasmonic field is dominated by a Bessel function $J_l(k_{sp}r)$ and possesses a phase factor exp($il\varphi$). This formula is the same as the result in [25] but under different excitation mechanism. In our case, due to polarization and geometrical symmetry, the SPP waves retain phase structure of the incident RPOV merely with a certain phase shift, and propagate inwardly from all azimuthal directions and interfere mutually to form the PV. For this symmetric structureless focusing system, the phase shift of $E_z$ in the process of passing through the metal film is determined by the thickness and dielectric constants of metal film and substrate, and it can be derived from the transmission coefficient $T_z^p$ of the p-polarized beam for substrate/Au/air three-layer configurations as $\delta = \arg(T_z^p) = 1.65\pi$.

Figure 2 shows the experimental near-field distribution of PV of $l$=2 recorded by NSOM (NT-MDT, NTEGRA Solaris) with an aluminum-coated fiber tip of 100 nm diameter aperture operated in collection mode, collecting at 60 nm above the Au film surface in order to verify the center dark spot due to phase singularity. As shown in Fig. 2(a), the interference of SPP waves propagating toward the geometrical center of the OV from all azimuthal directions form the center dark core and surrounding SPP standing waves. The measured intensity distribution along the white line in Fig.

2(a) is presented in Fig. 2(b). The surface plasmon interference period near the central region is measured to be 298±2 nm, in good agreement with theoretically predicted SPP half wavelength $\lambda_{sp}/2=\pi/\mathrm{Re}(k_{sp})$. The center of transverse profile of the experimental data does not reach the zero intensity is mainly due to the finite aperture size of the NSOM probe.

**Theoretical demonstration of formation and Poynting vector of PV.**

The instantaneous, amplitude and phase distributions of the $E_z$ component of the plasmonic field are numerically acquired and displayed in Figs. 3(a)-3(c) respectively by use of FullWAVE module of the commercial RSOFT software, a method based on FDTD methodology. The data in Fig. 3 are taken from the x-y plane 60 nm above the Au film, which is adequately within the decay length of SPP field (284 nm). The dashed black circle in Fig. 3(a) denotes the excitation boundary where the focused RPOV illuminates on the metal surface. Inside the circle, the interference of counter-propagating SPP results in standing waves in the radial direction and $2l$ patterns in the azimuthal direction as well as destructive interference induced dark core. The amplitude distribution $|E_z|$ has a ring of primary peak accompanied by concentric outer rings of diminishing intensity seen from Fig. 3(b). Period of the SPP pattern is evaluated as 298 nm which agrees well with half SPP wavelength and experimental results. From the simulation result as shown in Fig. 3(c), phase distributions show some peculiar behaviors in the interference region: the dark spot has a topological charge that represents $4\pi$ accumulated when the phase gradient is integrated around the center where the phase is undefined, and it has a $\pi$ phase shift across the node lines resulting in a series of intensity minima of the standing waves in the radial direction, and forms a close ring in the azimuthal direction. In each ring, the phase is divided into $l$ parts and monotonously increases from 0 to $2\pi$ in each part along the anticlockwise direction. Therefore, it can be inferred that illumination of highly focused RPOV on the planar metal surface induces the SPP fields with phase singularity. Different from the spiral phase originating from the geometric structure of the PVL, the direct transform of the spiral phase profile $\exp(il\varphi)$ from incident RPOV through high NA focusing system can make such a converging helical wavefront possible to synthesize PV around the converging center.

It is noteworthy that the transverse electric fields of PV, whose polarization determined by the radial ($E_r$) and azimuthal ($E_\varphi$) components is spatially varying, also possess phase singularity for integer topological charge $l\neq1$. For optical beams with space-variant polarization, the phase can be calculated using Pancharatnam's definition [34], that is to say, it is not the absolute phase of beam but the phase difference between observation points in different polarization states that determines the beam intensity profiles. Such phase difference can be represented by $\varphi_p = \arg\langle\vec{E}(r_1)|\vec{E}(r_2)\rangle$, where arg(x) gives the argument of the complex number x, $\langle\vec{E}(r_1)|\vec{E}(r_2)\rangle$ denotes an inner

product of two electric field vectors $\vec{E}(r_1)$ and $\vec{E}(r_2)$ at two different observation points on the wavefront. Considering two observation points located on the same circle around the beam axis, we can yield

$$\varphi_p = \arg\langle \vec{E}(r,0,z) | \vec{E}(r,\varphi,z) \rangle = (l-1)\varphi \tag{2}$$

Different from the phase factor exp($il\varphi$) carried by $E_z$ component of PV, it is seen that the phase of transverse electric field of PV is dependent on exp[$i(l-1)\varphi$], which reveals that in-plane electric field of PV has phase singularity at the beam center for $l\neq 1$. The phase distribution of transverse electric field of PV with topological charge $l$=2 acquired from FDTD simulation at the *x-y* plane located at 60 nm above the gold film is shown in Fig. 3 (d). Phase singularity is clearly seen at the beam center and the phase accumulation around the beam axis is $2\pi$, which agrees well with the theoretical prediction.

From the vector description of plasmonic fields of PV [25], the major component of Poynting vector is $S_{1\varphi}(r,\varphi,z) = \frac{\omega\varepsilon_0 l |E_{1z}|^2}{2rk_{sp}^2} J_l^2(k_{sp}r)\exp(-2k_z z)$. This dominant azimuthal component indicates that the energy flow of PV points to azimuthal direction and rotates along the beam axis. The longitudinal component $S_{1z}$ is zero because the evanescent field does not carry any energy in the decaying direction. We calculate the time-averaged Poynting vectors in the observation plane above the Au film by use of FDTD. The result is shown in Fig. 4 and agrees well with the theoretical predictions. It is emphasized that this movement of field energy is accompanied by an OAM of PV about the beam axis which will be experimentally demonstrated below.

Based on this explicit formula of Poynting vector, the focusing efficiency of PV in our excitation scheme can be calculated. It is defined as the ratio of optical power in the primary ring of the PV and the input power coupled into the objective lens, in the highly focusing configuration:

$$\eta = \frac{\int_0^\infty \left[\int_0^{R_{min}} S_{1\varphi}(r,\varphi,z)dr\right]dz}{\int_0^{2\pi}\left[\int_0^{R_{max}} S_{0z}(r,\varphi,z)rdr\right]d\varphi} = \frac{lk_0}{4\pi n k_z k_{SP}^2 w_0^2}\left|\frac{E_{1z}}{E_0}\right|^2 \frac{\int_0^{R_{min}}\left[J_l^2(k_{SP}r)/r\right]dr}{\int_0^{R_{max}} r^{2l+1}\exp(-2r^2)dr} \tag{3}$$

where $R_{min}$ and $R_{max}$ are radius of the first intensity minimum of PV and aperture size of the objective lens respectively, $w_0$ is the beam waist of OV beam profile represented as $u(r,z) = E_0 (\frac{r}{w_0})^l \exp(-\frac{r^2}{w_0^2})$, $S_{0z}$ denotes the *z*-component of time-averaged Poynting vector of incident RPOV. The optical power is obtained by integrating the Poynting vector component $S_{1\varphi}$ and $S_{0z}$ over the *r-z* and *x-y* plane for PV and input RPOV respectively. The focusing efficiency can numerically be evaluated through FDTD calculation. Here we give an example of charge $l$=5 using the parameters: $k_{sp}$=1.061$k_0$, $k_z$=0.355$k_0$, $w_0$~6μm, $R_{min}$~0.87μm, $R_{max}$~10μm. Substitution of $|E_{1z}/E_0|$~11.31 derived by FDTD into Eq.

(3), we yield $\eta \sim 3.38 \times 10^{-3}$. This value is higher than that of reported in Ref. [25] which is $3.13 \times 10^{-5}$ in the case of PV of $l=5$.

**Optical trapping metallic particles via OAM transfer from PV.**

To verify the PV having well-defined OAM conferred by the incident OV, we perform optical trapping experiment by transferring the OAM to absorptive particles and inducing them to rotate circumferentially in the azimuthal direction in the focal region of PV with different topological charges. Gold particles were chosen here to strengthen the transfer of OAM from PV to the particles because of their highly absorbing and scattering features. The diameters of the gold particles purchased from ALFA AESAR Co. were ranging from 0.8-1.5μm. In the experiment, the gold particles were diffused in water and injected into a homemade well. The use of metallic particles instead of dielectric spheres was to reduce the Brownian motion caused by thermal effect which can be detrimental in optical trapping, and to ensure that the particles adequately immersed in the highly localized evanescent SPP fields under the force of gravity and thus subsequent motions can be correlated to the SPP waves. If the focused PV possesses OAM, both polarization and topological charge change of incident OV can influence the motion and rotation rate of absorptive particles.

Figure 5 shows the typical results of near-field optical trapping experiment, in which a single gold particle was manipulated and rotated along the beam axis by PV of topological charges $l=2$ (top row) and $l=5$ (bottom row) via OAM transfer, respectively. A different working wavelength of 1064 nm (ND: YAG laser) was utilized here because of its prominently high optical power. Similar to the schematic setup in Fig. 1, we replaced the NSOM tip by an optical fiber illuminator (Thorlabs OSL 1-EC) to enable detection of the motion of gold particles by the CCD camera located at the particles' image plane of the objective lens. The beam splitter in Fig. 1 was replaced by a dichroic mirror (highly transmissive for 1064 nm and highly reflective for visible light), and a low-pass filter was positioned before the CCD to clear away the stray light of 1064 nm. The SPR angle is shifted to 63.4°, still within the range of the objective lens with a NA of 1.49. The angle was optimized to satisfy both the SPR excitation and observation of the particles via CCD. To make a quantitative comparison, the incident power measured by a power meter before entering the objective lens was adjusted to be 160 mW. We observed that the gold particle was trapped and rotated along the ring of PV several microns away from the central point. In order to unambiguously verify the OAM triggered by PV, the polarization was switched from radial to azimuthal by rotating one of the two half-wave plates by 45° in the beam path. In this case, no SPR was expected and the gold particle performed from rotation to stop. The rotational state of the particle was recovered when the polarization was turned back to the radial one. The orbital rotation speeds and radius of particles trapped with $l=2$ and $l=5$ OV were observed to be different in our experiment, and their product can be used to quantitatively characterize the OAM carried by PV, which demonstrates the proof-of-concept dynamic control capability. The average diameters of circular motion trajectories of the trapped gold particle with $l$=2 and $l$=5 are 0.88 μm and 1.55 μm and the average rotation speed of the particles trapped by $l$=2 and $l$=5 PV are 1.9 μm/s and 2.2 μm/s respectively. The product of rotation speed and orbital diameter of particles trapped by PV of $l$=5 was two times higher than that of $l$=2. The difference between this experimental value and theoretical prediction 2.5 (5/2) may be caused by the different transfer efficiency of OAM from incident OV with different topological charges to the surface plasmonic modes and different size of gold particles used in the two cases, which we haven't taken into consideration here. Therefore, it is feasible to dynamically switch PV by tuning the polarization and topological charge of incident OV.

**Discussion**

We have shown that plasmonic focusing into PV with arbitrary high order topological charges at any desired positions can be accomplished by means of highly focusing RPOV onto a planar metal surface. In such a highly focusing system, the incident RPOV can transfer its azimuthal phase and OAM to the plasmonic mode, which leads to confine and manipulate particles within the region of SPP evanescent field. The experimental results acquired from NSOM were confirmed by FDTD simulations and analytical calculations. The focusing efficiency in the highly focusing configuration is evaluated to be two orders higher than that of PVL. The OAM carried by PV is verified experimentally by rotating gold particles circumferentially around the beam center. The novel method for synthesis PV is simpler and more flexible than the slit pattern in spiral grooves and PVL that require the alignment between the centers of the radial polarizer and the PVL as well as fabrication of high-resolution nanostructures.

**Acknowledgements**

This work was partially supported by the National Natural Science Foundation of China under Grant No. (10974101, 61036013 and 61138003), Ministry of Science and Technology of China under Grant No.2009DFA52300 for China-Singapore collaborations and National Research Foundation of Singapore under Grant No. NRF-G-CRP 2007-01. XCY acknowledges the support given by Tianjin Municipal Science and Technology Commission under grant No. 11JCZDJC15200.

**Figure captions**

Fig. 1. Schematic diagram of experimental setup. The polarization states corresponding to the positions denoted by capital letters A, B, C, D in the optical path are sketchily given in the square with yellow background colour. The incident wavelength is 633 nm and spiral phase plate is utilized to produce the helical wavefront of OV. The spiral phase of topological charge $l$=1 is given in the lower right dashed box as an example. The red, green, blue and yellow arrows in upper right inset show the corresponding $E_z$ component with relative phase differences of 0, $\pi/2$, $\pi$ and $3\pi/2$, respectively.

Fig. 2. (a) Experimental NSOM image of the SPP excited by highly focused RPOV at 633 nm incident wavelength, (b) the intensity distribution along the white line in (a).

Fig. 3. Snapshot (a), amplitude (b) and phase (c) distributions of $E_z$ component as well as phase distribution of transverse electric field (d) of PV with topological charge $l$=2 in the *x-y* plane located at 60 nm above the Au film for RPOV excitation. The dashed black circle in (a) denotes the excitation boundary where the focused RPOV impinges on the metal surface. Singularity exists in the phase distribution at the center of the interference pattern.

Fig. 4. Time-averaged Poynting vector of PV of topological charge 2. The angular momentum flow points to the azimuthal direction.

Fig. 5. Successive frames of a video recording that show the rotation of one gold particle trapped by PV of $l$=2 (top row) and $l$=5 (bottom row). The circular arrow denotes the rotational direction and the dashed circles show the particle's orbital trajectory. The incident wavelength is 1064 nm.

Figure 1

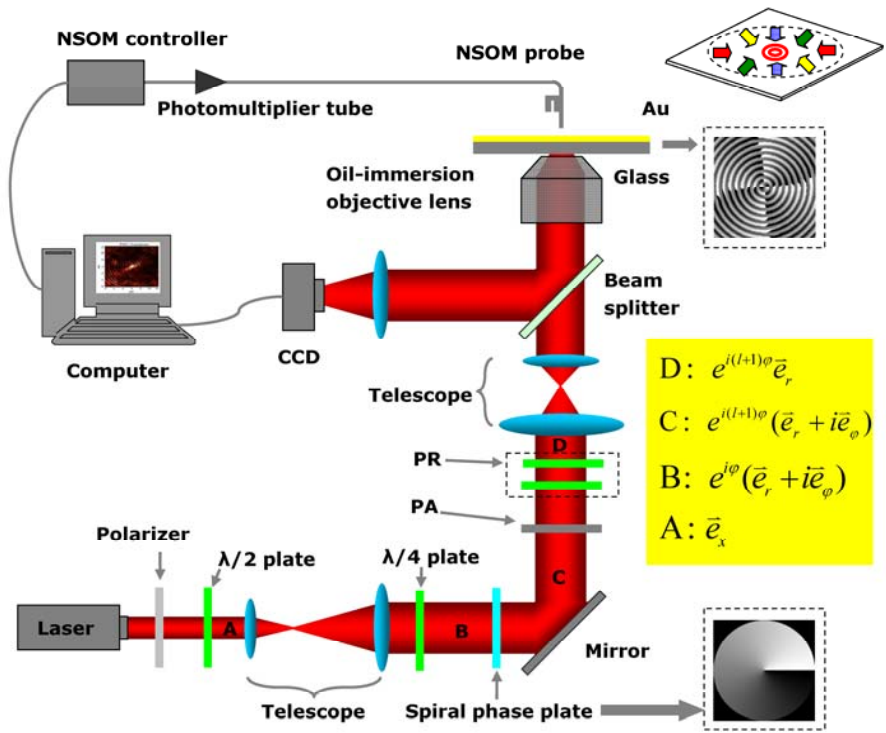

Figure 2

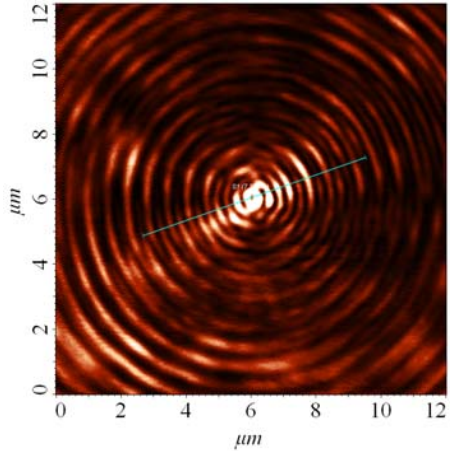 (a)
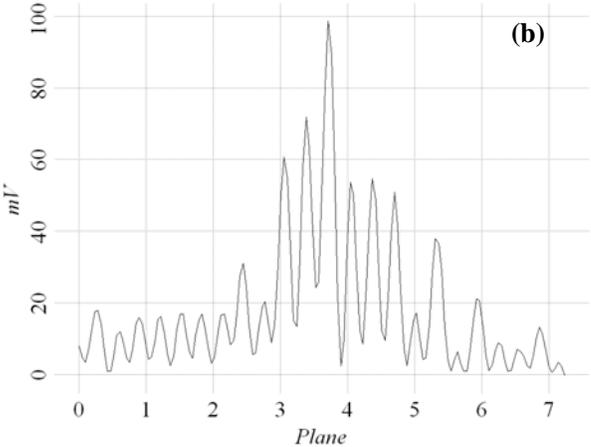 (b)

Figure 3

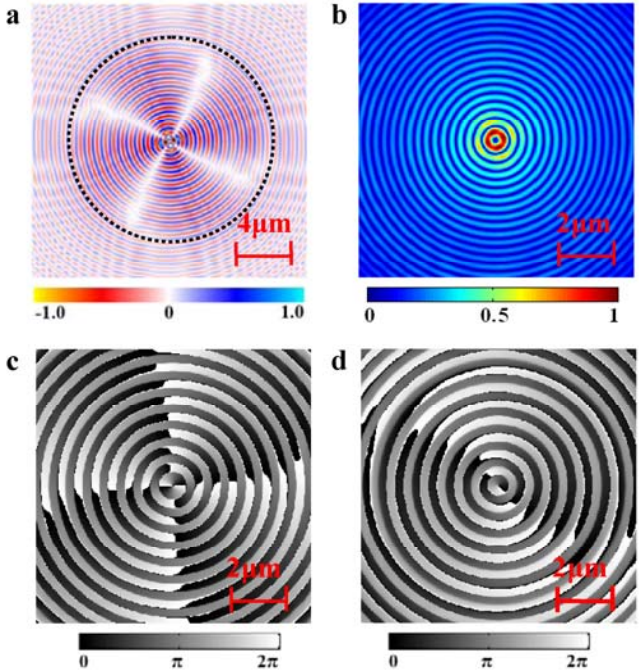

Figure 4

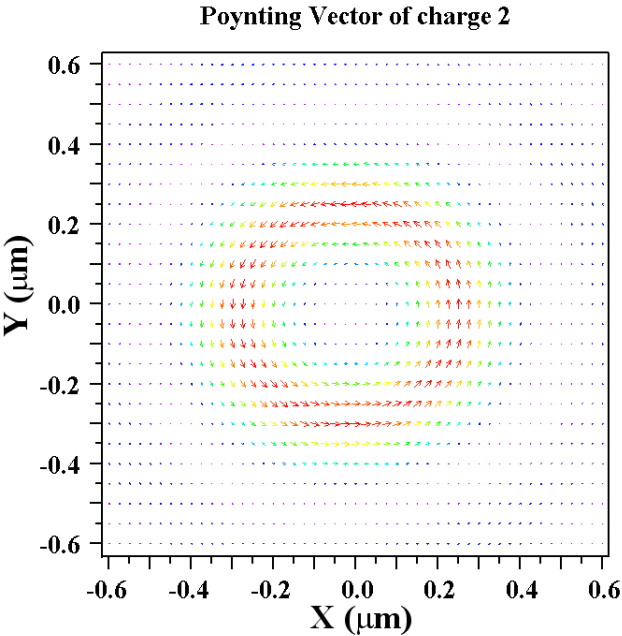

Figure 5

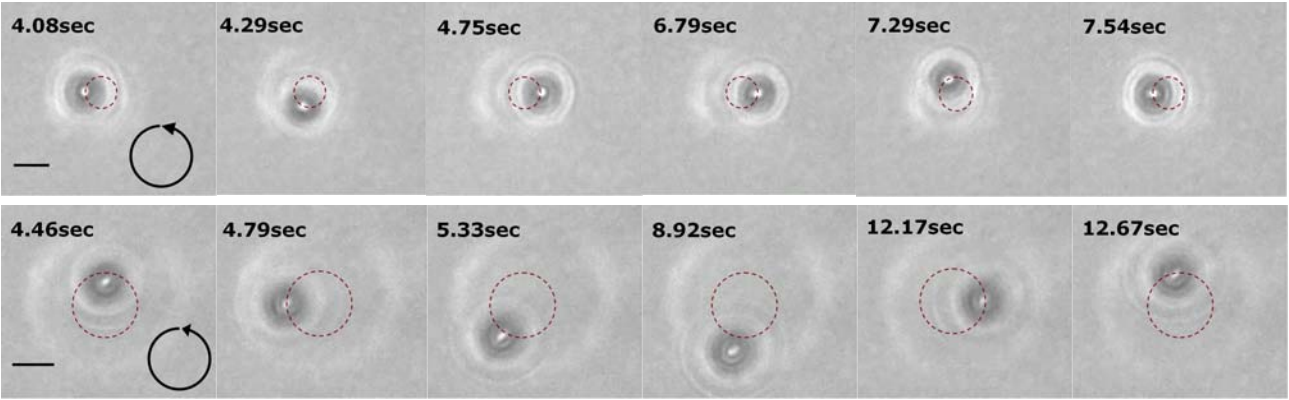